**Viscoelastic control of spatiotemporal order in bacterial active matter**

Song Liu* [1], Suraj Shankar* [2,3,4], M. Cristina Marchetti [5], Yilin Wu [1]

[1] *Department of Physics and Shenzhen Research Institute, The Chinese University of Hong Kong, Shatin, NT, Hong Kong, P.R. China.*

[2] *Department of Physics and Soft and Living Matter Program, Syracuse University, Syracuse, NY 13244, USA*

[3] *Kavli Institute for Theoretical Physics, University of California, Santa Barbara, CA 93106, USA*

[4] *Department of Physics, Harvard University, Cambridge, MA 02318, USA*

[5] *Department of Physics, University of California, Santa Barbara, CA 93106, USA*

* These authors contributed equally to this work.

**Text:**

Active matter consists of units that generate mechanical work by consuming energy [1]. Examples include living systems, such as assemblies of bacteria [2-5] and biological tissues [6,7], biopolymers driven by molecular motors [8-11], and suspensions of synthetic self-propelled particles [12-14]. A central question in the field is to understand and control the self-organization of active assemblies in space and time. Most active systems exhibit either spatial order mediated by interactions that coordinate the spatial structure and the motion of active agents[12,14,15] or the temporal synchronization of individual oscillatory dynamics [2]. The simultaneous control of spatial and temporal organization is more challenging and generally requires complex interactions, such as reaction-diffusion hierarchies [16] or genetically engineered cellular circuits [2]. Here, we report a novel and simple means to simultaneously control the spatial *and* temporal self-organization of bacterial active matter. By confining an active bacterial suspensions and manipulating a single macroscopic parameter, namely the viscoelasticity of the suspending fluid, we have found that the bacterial fluid first self-organizes in space into a millimeter-scale rotating vortex; then displays temporal organization as the giant vortex switches its global chirality periodically with tunable frequency, reminiscent of a torsional pendulum – a *self-driven* one. Combining experiments with an active matter model, we explain this striking behavior in terms of the interplay between active forcing and viscoelastic stress relaxation. Our findings advance the understanding of bacterial behavior in complex fluids, and demonstrate experimentally for the first time that rheological properties can be harnessed to control active matter flows[17,18]. When coupled to actuation systems, our millimeter-scale tunable, self-oscillating bacterial vortex may be used as a "clock generator" capable of providing timing signals for rhythmic locomotion of soft robots and for programmed microfluidic pumping[19], for example, via triggering the action of a shift register in soft-robotic logic devices[20].

Suspensions of swimming bacteria (bacterial active fluids) are important for bacterial dispersal and biofilm formation, and also offer a unique model system to study active matter self-organization [5,21]. Concentrated bacterial suspensions display intriguing rheological properties not seen in equilibrium, such as vanishing apparent viscosity at low shear [3,22]. Although in nature most bacteria swim in viscoelastic fluids, the role of viscoelasticity on bacterial dynamics, albeit considered theoretically [23,24], is largely unexplored experimentally. To examine whether fluid viscoelasticity modifies bacterial self-organization, we added purified genomic DNA from *E. coli* (~4.6 M base pairs, mol. wt. ~3.0x10$^9$ Dalton) to dense suspensions of *Escherichia coli* cells (0.8 μm in diameter, ~2-4 μm in length, swimming speed of ~20-40 μm/s) (Methods). *E. coli* DNA (hereinafter abbreviated as "DNA") was chosen because it has unusually high molecular weight and thus displays elastic response even at dilute concentrations [25]. This dense suspension of *E. coli* (~6×10$^{10}$ cells/mL) was deposited on the surface of agar gel (Methods) to form a disk-shaped liquid drop (~1.5 mm in diameter and ~20-30 μm in height at the center; Fig. 1a and Extended Data Figure 1a); the contact line of such a liquid drop is pinned to the substrate.

When the DNA concentration was dilute, the bacterial suspension displayed a disordered state with small-scale collective motion of cells (a few tens of μm) in the form of transient vortices or jets [5], known as bacterial or mesoscale "turbulence" [4]. When the DNA concentration reached >~50 ng/μL, we observed that the entire bacterial suspension drop rotated either clockwise (CW) or counterclockwise (CCW) at a constant angular speed of ~0.1-0.15 rad/s, forming a millimeter-scale unidirectional vortex (Video 1; Extended Data Figure 1b). To avoid confusion with the transient microscale vortices of bacterial turbulence, we refer to the millimeter-scale vortex observed here as "giant vortex". The collective velocity vectors obtained by particle image velocimetry (PIV) were well aligned in the giant vortex (Fig. 1b), and the azimuthally averaged tangential velocity increased with distance from the vortex center up to ~100-200 μm from the edge (Fig. 1c; Extended Data Figure 1c). The normalized mean vortical flow (i.e. tangential velocity averaged over the entire vortex; Methods) can also be used as an order parameter to characterize the vortex state, and it is indeed found to be near unity (Fig. 1d; Extended Data Figure 1d). On average, bacteria move coordinately along the advective drift in the giant vortex, reflecting the collective transport of the suspension, since the ambient fluid is dragged along by the cells [5]. Tracking of individual trajectories also reveals local diffusive behavior in a frame comoving with the vortex (Fig. 1e; Methods and Extended Data Figure 2a-b). Previously dense bacterial suspensions were reported to self-organize into stable vortices with coherent cell motion; such vortices had an upper size limit of ~100 μm, beyond which the collective motion became turbulent [21]. In stark contrast, the giant vortex we observed here is one order of magnitude greater in size, showing that additive DNA facilitates large-scale spatial ordering of bacterial active fluids.

Strikingly, when the concentration of DNA was increased further (>~300 ng/µL), the unidirectional giant vortex transitions into an oscillatory state, in which the global rotational chirality switches between CW and CCW with a well-defined period (Fig. 2a,b; Video 2-4). Meanwhile individual bacteria still displayed local diffusive behavior in a frame comoving with the vortex (Extended Data Figure 2c). Numerical solution of the continuum active matter model described below also reproduces the transition from coherent to oscillatory vortical flows (Fig. 2c,d and Video 5-6). The oscillation dynamics of the giant vortex is clearly seen in the temporal evolution of the mean vortical flow (Fig. 2b) and of the tangential velocity profile along the radial direction (Fig. 2e,f). The period of chirality switching is accurate with <~20% error, as revealed by Fourier spectrum analysis (Extended Data Figure 1e-g). Interestingly, the period of chirality switching can be tuned by DNA concentration; it increased from ~10 s to ~50 s when the DNA concentration was increased from ~300 ng/µL to ~800 ng/µL. Another important feature of the oscillatory giant vortex is that it acts like a relaxation oscillator [26]: the system transits quickly towards tangential (or angular) velocity extrema and progresses slowly away from the extrema, as manifested by the asymmetric shape of the velocity oscillation in Fig. 2f and by the parallelogram-like trajectory in the phase space of angular velocity and rotational angle (Fig. 2g). By contrast, the phase space trajectory of a sinusoidal oscillation would have an elliptical shape. We further examined the dynamics of global chirality switching of the giant vortex. At the initial stage of switching, a local vortex with opposite chirality tended to emerge near the periphery of the giant vortex (Fig. 3a). As the local vortex subsequently grew in size, a clear boundary with prominent local vorticity (referred to as the "switching front") was formed (Fig. 3b; Methods). The space-time plot in Fig. 3c clearly shows the propagation of the switching front (Fig. 3d).

Cell density is an important control parameter for bacterial collective motion and self-organization [27]. We found that there exists a critical cell density of ~3×10^10 cells/mL below which we could not observe the robust unidirectional giant vortex, nor the oscillatory one. At any cell density above this critical value, there exist two threshold DNA concentrations marking the onset of spatial order (unidirectional giant vortex) and temporal order (periodic switching of global rotational chirality of the giant vortex), denoted as $d_1$ and $d_2$ respectively. Although unidirectional giant vortex could be observed occasionally at DNA concentrations ~20-50 ng/µL (Extended Data Figure 3; Methods), it only developed robustly at ~50 ng/µL for all cell densities, suggesting that $d_1$ remains fairly constant and can be approximately taken as ~ 50 ng/µL. We found that $d_2$ decreases from ~400 ng/µL to ~60 ng/µL as cell density increases from $4\times10^{10}$ cells/mL to $8\times10^{10}$ cells/mL (Fig. 4a). Moreover, as shown in the phase diagram of Fig. 4a, the chirality-switching frequency of the oscillatory giant vortex decreases with increasing cell density or DNA concentration over a 6-fold tunable range between ~0.02 Hz and ~0.12 Hz; the tunable range is primarily controlled by DNA concentration and to a lesser extent by cell density. The amplitude of oscillations increases weakly with cell density and DNA concentration (Extended Data Figure 4). These results demonstrate that, given sufficiently high cell density, additive DNA polymers regulate both spatial and temporal self-organization of bacterial active fluids.

To understand the mechanism underlying the self-organized oscillations, we model the bacterial suspension as an active polar bacterial fluid coupled to a viscoelastic solvent [11,18,28]. The local bacterial orientation is described by a polarization vector $\boldsymbol{p}$ coupled to the fluid flow velocity $\mathbf{v}$ and the elastic stress $\boldsymbol{\sigma}^{el} = 2G'\boldsymbol{\varepsilon}$ of the DNA polymer, with $G'$ the polymer storage modulus and $\boldsymbol{\varepsilon}$ the strain. Assuming both the density of the suspension and the bacteria concentration to be constant, a minimal description of the active liquid crystal dynamics coupled to polymeric stress is given by

$$D_t\boldsymbol{p} + \boldsymbol{\Omega}\cdot\boldsymbol{p} = \frac{1}{\gamma}\boldsymbol{h} + \lambda\boldsymbol{S}\cdot\boldsymbol{p} - \frac{1}{\tau_R G'}\boldsymbol{\sigma}^{el}\cdot\boldsymbol{p} \qquad [1]$$

$$D_t\boldsymbol{\sigma}^{el} + \boldsymbol{\Omega}\cdot\boldsymbol{\sigma}^{el} - \boldsymbol{\sigma}^{el}\cdot\boldsymbol{\Omega} = -\frac{1}{\tau_p}\boldsymbol{\sigma}^{el} + 2G'\boldsymbol{S} \qquad [2]$$

with $D_t = \partial_t + \mathbf{v}\cdot\nabla$, $\Omega_{ij} = (\partial_i v_j - \partial_j v_i)/2$ the vorticity tensor and $S_{ij} = (\partial_i v_j + \partial_j v_i)/2$ the rate of strain tensor. The molecular field $\boldsymbol{h} = [a(c - c_0) - b|\boldsymbol{p}|^2]\boldsymbol{p} + K\nabla^2\boldsymbol{p}$, with $a, b > 0$ and a single elastic constant $K$, yields a transition from a disordered ($|\boldsymbol{p}| = 0$) to a polar ordered state with $|\boldsymbol{p}| = p_0 = \sqrt{a(c - c_0)/b}$ at cell density $c = c_0$. Vorticity and strain can rotate bacterial alignment $\boldsymbol{p}$ (the flow alignment parameter $\lambda > 1$ for elongated swimmers), with relaxation controlled by the rotational viscosity $\gamma$. The last term in Eq [1] is the simplest strain-polarization coupling, with $\tau_R$ an orientational relaxation time that controls the alignment of bacterial polarization to polymer strain, similar to that in passive liquid crystal elastomers [29]. The DNA is modeled as a standard elastic medium, with Maxwell relaxation time $\tau_p = \eta/G'$ and $\eta$ the shear viscosity. The flow velocity is determined by the Stokes equation that imposes force balance, $\Gamma(\mathbf{v} - v_0\mathbf{p}) = \nabla \cdot (\boldsymbol{\sigma}^{el} + \boldsymbol{\sigma}^a) - \nabla\Pi$, with $\Gamma$ the substrate friction, $v_0$ the bacteria swimming speed, and $\Pi$ the pressure required to enforce incompressibility. In the experiments the thickness of the bacterial drop is much smaller than its lateral size, indicating that friction dominates over viscous stresses. The active stress is $\boldsymbol{\sigma}^a = \alpha\mathbf{pp}$, with $a < 0$ for pushers such as *E. coli*. As the active stress is proportional to the average force dipole exerted by the swimmers, we expect $|\alpha| \sim c$. The dynamics is controlled by three competing time scales: the Maxwell relaxation time $\tau_p$, the stress alignment time $\tau_R$, and the active shearing time $\tau_a = \Gamma \ell_a^2/|\alpha|$, with $\ell_a \sim \sqrt{K/|\alpha|}$ a characteristic length scale (as in active nematics [30]). Numerical solution of the continuum model (details described in the Supplementary Information Sec. III) reproduces the transition from a global vortex state to an oscillatory state with periodic flow reversal (Fig. 2c,d; Supplementary Information Fig. 2, Fig. 3; Video 5,6) and shows that the transition is controlled by the interplay of these three time scales. An analytical analysis of steady states and their stability (see Supplementary Information Secs. IV and V) confirms the numerics and yields stability boundaries summarized in a phase diagram in Fig. 4b (note that only the high concentration part inscribed by the black box is relevant here). Briefly, increasing $\tau_p$ (that grows with DNA concentration) at fixed cell density, the system first transits from the turbulent state to polar laminar flow at $\tau_p = \tau_I \sim \tau_R$ via suppression of the splay instability [1,31], corresponding to a unidirectional giant vortex at a DNA concentration $d_1$ (~50 ng/μL) essentially independent of cell density in Fig. 4a, then to an oscillatory state at $\tau_p = \tau_{II} \sim \tau_a \sim K/\alpha^2$ with an oscillation frequency $\omega \sim |\alpha|/\sqrt{\tau_R} \sim 1/\sqrt{\tau_p}$ at threshold, corresponding to the oscillatory giant vortex at a DNA concentration $d_2$ that decreases with increasing cell density in Fig. 4a. The numerics also suggests that, although in the oscillatory state the bacterial polarization only exhibits small transverse oscillations about its mean direction while the velocity reverses, these transverse polarization fluctuations are responsible for the instability of the giant vortex (see Supplementary Information Fig. 2). This observation allows us to map the nonlinear dynamics onto the FitzHugh-Nagumo model[32], a well-known excitable relaxation oscillator (Extended Data Figure 5 and Supplementary Information Sec. VII), and show that the transition to spontaneous oscillations at $\tau_p \sim \tau_a$ is via a Hopf bifurcation, indicating a possible mechanism for the experimental results. Further details of the calculation and simulations can be found in the Supplementary Information.

In the experiment, $\tau_p$ indeed approaches $\tau_a$ when the system transits to oscillations (Extended Data Figure 6). Consistent with the experimental observations, the transition to oscillations occurs at $\tau_{II} \sim K/\alpha^2$, which decreases with increasing bacterial concentration (activity) (Fig. 4a); and the oscillation frequency $\omega$ decreases with addition of DNA (Fig. 4c; Methods), since $\tau_p$ increases with DNA concentration (Fig. 4d) and $\tau_R$ is expected to behave similarly. This feature is also corroborated by the fully nonlinear simulations (Supplementary Information Fig. 3c-e). On the other hand, $\omega$ also increases with activity, hence with bacterial concentration. This is at odds with experiments that find that the oscillation frequency *decreases* with bacteria concentration, but is a generic feature of active matter models that display relaxation oscillation with [18,28] or without [17,33,34] added polymer. It is unclear at present how this discrepancy may be resolved, but it suggests that the effects of nonlinear viscoelasticity in active fluids deserve more attention.

Despite the success of our theoretical model in explaining the essential phenomena, several open questions remain. First, the model suggests that transverse polarization fluctuations drive the switching of the flow, but more work is needed to firmly establish the connection between the change in sign of the splay and flow switching. Second, shear bands are observed in the model (Video 5) and only rarely in the experiments. Understanding the role of boundary conditions on shear banding will require extensive simulations of the model. Third, our numerical data is not sufficient to conclusively confirm non-sinusoidal oscillations as shown by experiments (Fig. 2d). Relaxation oscillators in general exhibit sinusoidal (i.e., harmonic) oscillations close to the Hopf bifurcation, with the oscillation waveform becoming strongly anharmonic far from threshold. Hence the approximate sinusoidal nature of the oscillations in the simulation may be a result of proximity to the Hopf bifurcation. Finally, the numerical resolution is too low to resolve the switching dynamics of giant vortices. In the experiments there may be other time scales that slow down the switching dynamics. Addressing this will require augmenting the model and is left for future work.

Our model suggests that the ultra-long relaxation times ($\tau_p$) of high molecular weight DNA [25] is key to the spatial-temporal order we revealed. Indeed, we observed the formation of giant vortices with other types of high molecular weight DNA with $\tau_p$ on the order of seconds, but not with viscoelastic polymers with $\tau_p$ at the millisecond scale (Methods; Extended Data Figure 7). Moreover, increasing medium viscosity tends to reduce cell speed and does not promote the formation of giant vortices (Extended Data Figure 8). In addition, the storage modulus ($G'$) of the polymer must be sufficiently large, such that the resulting elastic stress can affect the collective motion pattern of the bacteria [35]. DNA viscoelasticity contributes to $G'$, but we found that bacterial suspensions *without* additive polymers also display elasticity (~0.01 Pa) above cell density of about $4 \times 10^{10}$ cells/mL (Fig. 4e; measured on the scale of ~100 µm), which coincides with the critical cell density required for the onset of unidirectional giant vortex. Finally, we stress that spatial confinement is essential, as we could not observe the giant vortex in bacterial suspension drops with a diameter above ~3.3 mm. Nonetheless, by varying the size of suspension drops from ~1 mm to ~2.5 mm, we found that the threshold for the transition from bacterial turbulence to the giant vortex ($d_1$) is largely insensitive to confinement size (Methods; Extended Data Figure 10).

Taken together, our results demonstrate that tuning fluid viscoelasticity provides a simple means for manipulating the self-organization of bacterial active matter in space and time. Bacteria in biofilms and animal gastrointestinal tracts often swim in viscoelastic fluids abundant with long-chain polymers, including extracellular DNA [36]. Our findings suggest that, above a threshold bacterial density, the viscoelasticity of the environment may modify the collective motion patterns of bacteria, thereby influencing the dispersal of biofilms and the translocation of gut microbiome. We have developed a minimal active matter model that explains our findings as arising from the interplay between polymer viscoelastic relaxation and the rate of active forcing. Our work may shed light on the role of environment viscoelasticity in other active systems, such as cytoskeletal fluids [9,10] and active gels [11]. It may also pave the way to the development of a new class of adaptive self-driven devices and materials that exploits the feedback between activity and viscoelasticity.

## Methods

No statistical methods were used to predetermine sample size.

Two *E. coli* stains were used: HCB1737 (a derivative of E.coli AW405; from Howard Berg, Harvard University, Cambridge, MA) and HCB1737 GFP (HCB1737 with constitutive expression of green fluorescent protein encoded on the plasmid pAM06-tet [37] from Arnab Mukherjee and Charles M. Schroeder, University of Illinois at Urbana-Champaign). *E. coli* genome DNA was purified with Genomic DNA Purification Kit from Promega (Cat. No. A1120), following the protocol provided by manufacturer. DNA concentration was measured by Nanodrop Spectrophotometer (Thermofisher). Rheological measurements of bacteria suspension were performed in a rheometer (Anton Paar Physica MCR 301) or by microrehology measurement (Extended Data Figure 9b) [38,39,40].

Collective motion of bacterial suspension was observed in phase contrast with a 4× objective (Nikon Plan Fluor 4×, numerical aperture 0.13, working distance 16.5 mm) mounted on an inverted microscope (Nikon Eclipse Ti). Recordings were made with an sCMOS camera (Andor Zyla 4.2) at 30 fps. In all experiments the Petri dishes were covered with a lid to prevent evaporation and air convection, and the sample temperature was maintained at 30 °C using a custom-built temperature control system installed on microscope stage.

The velocity field of bacterial collective motion $v(r, t)$ was obtained by performing particle image velocimetry (PIV) on phase contrast microscopy images using an open-source package MatPIV 1.6.1 written by J. Kristian Sveen (http://folk.uio.no/jks/matpiv/index2.html). The vortex order parameter, i.e. normalized mean vortical flow ($P$), is defined as $P(t) = \langle v(r, t) \cdot e_\theta / |v(r, t)| \rangle_{r,\theta}$, where $e_\theta$ is the unit vector along tangential direction (in the polar coordinate system whose origin is located at the center of the suspension drop) and the angular brackets indicate averaging over polar coordinates $r$ and $\theta$. $P$ being equal to +1 (or -1) indicates perfectly ordered CCW (or CW) vortex. Unidirectional giant vortexes typically have $|P| > 0.6$ averaged over time.

Single cells were tracked for at least 10 s in fluorescent images using the MTrackJ plugin developed for ImageJ.  The background bacterial collective velocity field was computed by performing PIV analysis on phase contrast images obtained simultaneously with the fluorescent images.  To compute the drift-corrected mean square displacement (MSD) of single cells, the local advective drift was taken as the average of bacterial collective velocity in a circular region with a radius of 15 μm and centered at the tracked bacterium, and then the obtained local advective drift was subtracted from the velocity of the cell.  The resulted drift-corrected single-cell velocity was integrated over time to find the drift-corrected displacement, which was further used to calculate the MSD.

**Code availability.**  The custom codes used in this study are available from the corresponding author upon request.

**References**

1   Marchetti, M. C. *et al.* Hydrodynamics of soft active matter. *Reviews of Modern Physics* **85**, 1143-1189 (2013).

2   Danino, T., Mondragon-Palomino, O., Tsimring, L. & Hasty, J. A synchronized quorum of genetic clocks. *Nature* **463**, 326-330, doi:http://www.nature.com/nature/journal/v463/n7279/suppinfo/nature08753_S1.html (2010).

3   Sokolov, A. & Aranson, I. S. Physical properties of collective motion in suspensions of bacteria. *Phys Rev Lett* **109**, 14 (2012).

4   Wensink, H. H. *et al.* Meso-scale turbulence in living fluids. *Proceedings of the National Academy of Sciences* **109**, 14308-14313, doi:10.1073/pnas.1202032109 (2012).

5   Chen, C., Liu, S., Shi, X. Q., Chate, H. & Wu, Y. Weak synchronization and large-scale collective oscillation in dense bacterial suspensions. *Nature* **542**, 210-214, doi:10.1038/nature20817 (2017).

6   Saw, T. B. *et al.* Topological defects in epithelia govern cell death and extrusion. *Nature* **544**, 212, doi:10.1038/nature21718(2017).

7   Kawaguchi, K., Kageyama, R. & Sano, M. Topological defects control collective dynamics in neural progenitor cell cultures. *Nature* **545**, 327, doi:10.1038/nature22321 (2017).

8   Keber, F. C. *et al.* Topology and dynamics of active nematic vesicles. *Science* **345**, 1135-1139, doi:10.1126/science.1254784 (2014).

9   Wu, K.-T. *et al.* Transition from turbulent to coherent flows in confined three-dimensional active fluids. *Science* **355**, doi:10.1126/science.aal1979 (2017).

10  Huber, L., Suzuki, R., Krüger, T., Frey, E. & Bausch, A. R. Emergence of coexisting ordered states in active matter systems. *Science*, doi:10.1126/science.aao5434 (2018).

11  Prost, J., Jülicher, F. & Joanny, J. F. Active gel physics. *Nature Physics* **11**, 111, doi:10.1038/nphys3224 (2015).

12  Palacci, J., Sacanna, S., Steinberg, A. P., Pine, D. J. & Chaikin, P. M. Living Crystals of Light-Activated Colloidal Surfers. *Science* **339**, 936-940, doi:10.1126/science.1230020 (2013).

13  Bricard, A., Caussin, J.-B., Desreumaux, N., Dauchot, O. & Bartolo, D. Emergence of macroscopic directed motion in populations of motile colloids. *Nature* **503**, 95-98, doi:10.1038/nature12673 (2013).

14  Yan, J. *et al.* Reconfiguring active particles by electrostatic imbalance. *Nature Materials* **15**, 1095, doi:10.1038/nmat4696 (2016).

15  Karig, D. *et al.* Stochastic Turing patterns in a synthetic bacterial population. *Proceedings of the National Academy of Sciences*, doi:10.1073/pnas.1720770115 (2018).

16  Vicker, M. G. Eukaryotic Cell Locomotion Depends on the Propagation of Self-Organized Reaction–Diffusion Waves and Oscillations of Actin Filament Assembly. *Experimental Cell Research* **275**, 54-66, doi:https://doi.org/10.1006/excr.2001.5466 (2002).

17  Giomi, L., Mahadevan, L., Chakraborty, B. & Hagan, M. F. Banding, excitability and chaos in active nematic suspensions. *Nonlinearity* **25**, 2245 (2012).

18  Hemingway, E. J. *et al.* Active Viscoelastic Matter: From Bacterial Drag Reduction to Turbulent Solids. *Physical Review Letters* **114**, 098302, doi:10.1103/PhysRevLett.114.098302 (2015).

19  Wehner, M. *et al.* An integrated design and fabrication strategy for entirely soft, autonomous robots. *Nature* **536**, 451, doi:10.1038/nature19100 (2016).

20  Preston, D. J. *et al.* Digital logic for soft devices. **116**, 7750-7759, doi:10.1073/pnas.1820672116 (2019).

21  Wioland, H., Woodhouse, F. G., Dunkel, J., Kessler, J. O. & Goldstein, R. E. Confinement Stabilizes a Bacterial Suspension into a Spiral Vortex. *Physical Review Letters* **110**, 268102 (2013).

22  López, H. M., Gachelin, J., Douarche, C., Auradou, H. & Clément, E. Turning Bacteria Suspensions into Superfluids. *Physical Review Letters* **115**, 028301 (2015).

23  Bozorgi, Y. & Underhill, P. T. Effects of elasticity on the nonlinear collective dynamics of self-propelled particles. *Journal of Non-Newtonian Fluid Mechanics* **214**, 69-77, doi:https://doi.org/10.1016/j.jnnfm.2014.09.016 (2014).

24  Li, G. & Ardekani, A. M. Collective Motion of Microorganisms in a Viscoelastic Fluid. *Physical Review Letters* **117**, 118001, doi:10.1103/PhysRevLett.117.118001 (2016).

25  Liu, Y., Y., J. & V., S. Concentration dependence of the longest relaxation times of dilute and semi-dilute polymer solutions. *Journal of Rheology* **53**, 1069-1085, doi:10.1122/1.3160734 (2009).

26  Ginoux, J. & Letellier, C. Van der Pol and the history of relaxation oscillations: Toward the emergence of a concept. *Chaos: An Interdisciplinary Journal of Nonlinear Science* **22**, 023120, doi:10.1063/1.3670008 (2012).

27  Sokolov, A., Aranson, I. S., Kessler, J. O. & Goldstein, R. E. Concentration Dependence of the Collective Dynamics of Swimming Bacteria. *Physical Review Letters* **98**, 158102 (2007).

28  Hemingway, E. J., Cates, M. E. & Fielding, S. M. Viscoelastic and elastomeric active matter: Linear instability and nonlinear dynamics. *Phys Rev E* **93**, 032702, doi:10.1103/PhysRevE.93.032702 (2016).

29  Warner, M. & Terentjev, E. M. *Liquid Crystal Elastomers.* (OUP Oxford, 2007).

30  Doostmohammadi, A., Ignés-Mullol, J., Yeomans, J. M. & Sagués, F. Active nematics. *Nature Communications* **9**, 3246, doi:10.1038/s41467-018-05666-8 (2018).

31  Aditi Simha, R. & Ramaswamy, S. Hydrodynamic Fluctuations and Instabilities in Ordered Suspensions of Self-Propelled Particles. *Physical Review Letters* **89**, 058101, doi:10.1103/PhysRevLett.89.058101 (2002).

32  Murray, J. D. *Mathematical Biology: I. An Introduction.* (Springer New York, 2007).

33  Giomi, L., Mahadevan, L., Chakraborty, B. & Hagan, M. F. Excitable Patterns in Active Nematics. *Physical Review Letters* **106**, 218101, doi:10.1103/PhysRevLett.106.218101 (2011).

34  Woodhouse, F. G. & Goldstein, R. E. Spontaneous Circulation of Confined Active Suspensions. *Physical Review Letters* **109**, 168105, doi:10.1103/PhysRevLett.109.168105 (2012).

35  Benzi, R. & Ching, E. S. C. Polymers in Fluid Flows. *Annual Review of Condensed Matter Physics* **9**, 163-181, doi:10.1146/annurev-conmatphys-033117-053913 (2018).

36  Whitchurch, C. B., Tolker-Nielsen, T., Ragas, P. C. & Mattick, J. S. Extracellular DNA Required for Bacterial Biofilm Formation. *Science* **295**, 1487-1487, doi:10.1126/science.295.5559.1487 (2002).

Additional references in Methods and Extended Data Figures

37  Mukherjee, A., Walker, J., Weyant, K. B. & Schroeder, C. M. Characterization of Flavin-Based Fluorescent Proteins: An Emerging Class of Fluorescent Reporters. *PLOS ONE* **8**, e64753, doi:10.1371/journal.pone.0064753 (2013).

38  Mason, T. G., Ganesan, K., van Zanten, J. H., Wirtz, D. & Kuo, S. C. Particle Tracking Microrheology of Complex Fluids. *Physical Review Letters* **79**, 3282-3285, doi:10.1103/PhysRevLett.79.3282 (1997).

39  Zhu, X., B., K. & J.R.C., v. d. M. Viscoelasticity of entangled λ-phage DNA solutions. *The Journal of Chemical Physics* **129**, 185103, doi:10.1063/1.3009249 (2008).

40  Kundukad, B. & van der Maarel, J. R. C. Control of the Flow Properties of DNA by Topoisomerase II and Its Targeting Inhibitor. *Biophysical Journal* **99**, 1906-1915, doi:https://doi.org/10.1016/j.bpj.2010.07.013 (2010).

41  Brochard, F. Viscosities of dilute polymer solutions in nematic liquids. *Journal of Polymer Science: Polymer Physics Edition* **17**, 1367-1374, doi:10.1002/pol.1979.180170808 (1979).

**Supplementary Materials**, including Supplementary Information with details of the active matter model and Supplementary Videos, is available in the online version of the paper.

**Data availability.** The data supporting the findings of this study are included within the paper and its Supplementary Materials.

**Code availability.** The custom codes used in this study are available from the corresponding author upon request.

**Acknowledgements**. We thank Ye Li and Wenlong Zuo for building the image acquisition and microscope stage temperature control systems, Howard C. Berg (Harvard University) for providing the bacterial strains, Arnab Mukherjee and Charles M. Schroeder (UIUC) for providing the pAM06-tet plasmid, and Lei Xu (CUHK) for assistance with bulk rheology measurement. We thank Emily S.C. Ching (CUHK), Keqing Xia (CUHK) and To Ngai (CUHK) for helpful discussions and comments. This work was supported by the National Natural Science Foundation of China (NSFC No. 31971182, to Y.W.), the Research Grants Council of Hong Kong SAR (RGC Ref. No. 14303918 and CUHK Direct Grants; to Y.W.), the US National Science Foundation Grant DMR-1609208 (to M.C.M and S.S) and KITP under Grant No. PHY-1748958. S.S. is supported by the Harvard Society of Fellows. M.C.M and S.S thank the KITP for hospitality in the course of this work.

**Author Contributions**: S.L. discovered the phenomena, designed the study, performed experiments, analyzed and interpreted the data. S.S. and M.C.M developed the active matter model, analyzed and interpreted the data. Y.W. conceived the project, designed the study, analyzed and interpreted the data. Y.W. wrote the first draft and all authors contributed to the revision of the paper.

**Author Information**: Reprints and permissions information is available at www.nature.com/reprints. The authors declare no competing financial interests. Correspondence and requests for materials should be addressed to Y.W. (ylwu@cuhk.edu.hk).

**Figures**

Figure 1

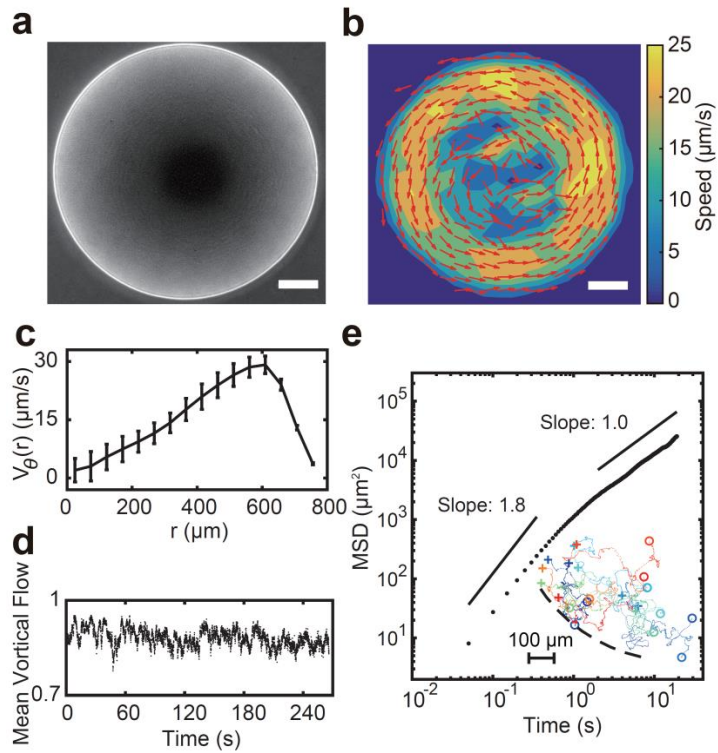

**Figure 1. Unidirectional giant vortex.** (**a,b**) Phase contrast image and instantaneous velocity field of a unidirectional giant vortex. Arrows and colormap in **b** represent collective velocity direction and magnitude, respectively (Methods). DNA concentration, 200 ng/μL. Scale bars, 250 μm. Also see Video 1. (**c**) Time- and azimuthally averaged tangential velocity $v_\theta(r)$ of the giant vortex in **b** plotted against radial position. Error bars represent standard deviation (N=1000 successive frames). (**d**) Normalized mean vortical flow of the giant vortex in **b** (Methods). (**e**) Drift-corrected mean square displacement (MSD) of single cells in a giant vortex (Methods; Extended Data Figure 2). Inset: Trajectories of 11 representative cells (+: starting point; ○: ending point; dashed line: edge of suspension drop).

Figure 2

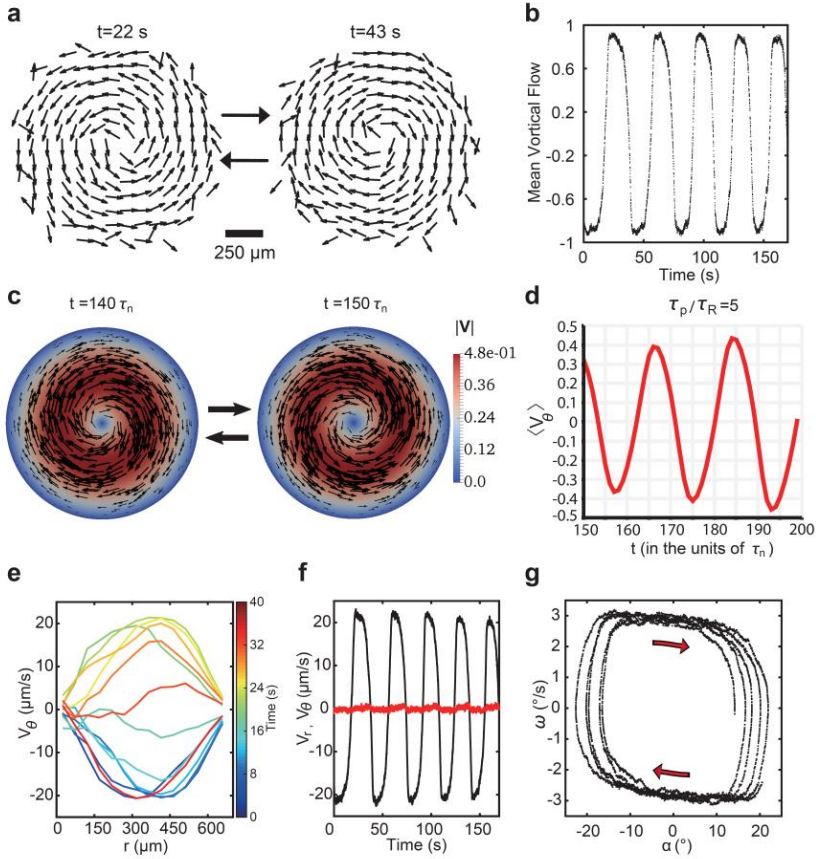

**Figure 2.  Oscillatory giant vortex.** (**a**) Two snapshots of the velocity direction field of an oscillatory giant vortex switching its global rotational chirality every ~35 s.  DNA concentration, 800 ng/μL.  (**b**) Periodic chirality switching indicated by the oscillation of normalized mean vortical flow (positive: CCW; negative: CW).  (**c**) Two simulation snapshots showing periodic reversal of flow velocity.  $\tau_n$ denotes the natural relaxation time of the bacteria orientation (Supplementary Information Sec. I).  Arrows and colormap represent velocity direction and magnitude, respectively.  (**d**) Time trace of the mean vortical flow $\langle V_\theta \rangle$ (Methods) associated with simulation results in **c**.  (**e**) Temporal evolution of azimuthally averaged tangential velocity $v_\theta$ during chirality-switching.  Colormap indicates time.  (**f**) Time trace of azimuthally averaged tangential ($v_\theta$; black) and radial ($v_r$; red) velocity computed near half radius of the giant vortex (390 μm ≤ $r$ ≤ 440 μm).  (**g**) Phase space trajectory of the oscillatory giant vortex in the plane of angular velocity $\omega$ and rotational angle $\alpha$.  $\omega$ is computed as $v_\theta/r$ in **f**, and $\alpha$ is computed by integrating $\omega$ over time.  Also see Videos 2-6.

Figure 3

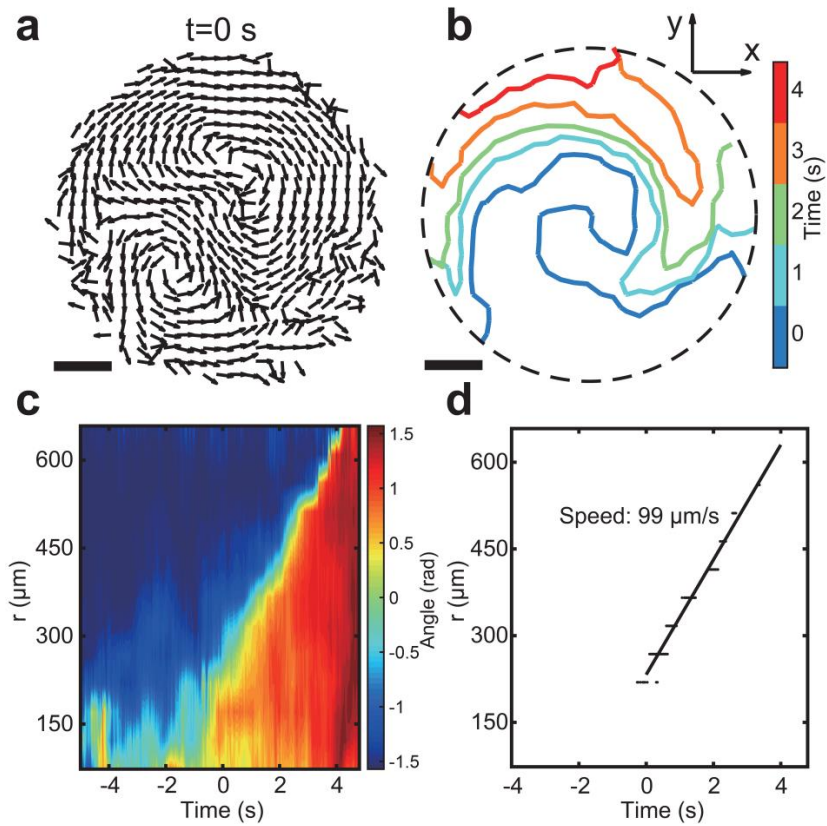

**Figure 3. Dynamics of chirality switching process in oscillatory giant vortex.** (**a**) Direction field of collective velocity at the initial stage of a chirality-switching event showing the emergence of a local vortex with opposite chirality (Video 3). (**b**) Propagation of the "switching front" (colored lines; Methods). Colormap indicates time. Scale bars in **a**, **b**, 250 µm. (**c**) Space-time plot (kymograph) of the direction of collective velocities (i.e. relative angle between collective velocity and local radial orientation) (Methods). (**d**) The propagation speed of chirality switching front in **c** is computed by fitting the scattered points of switching front.

Figure 4

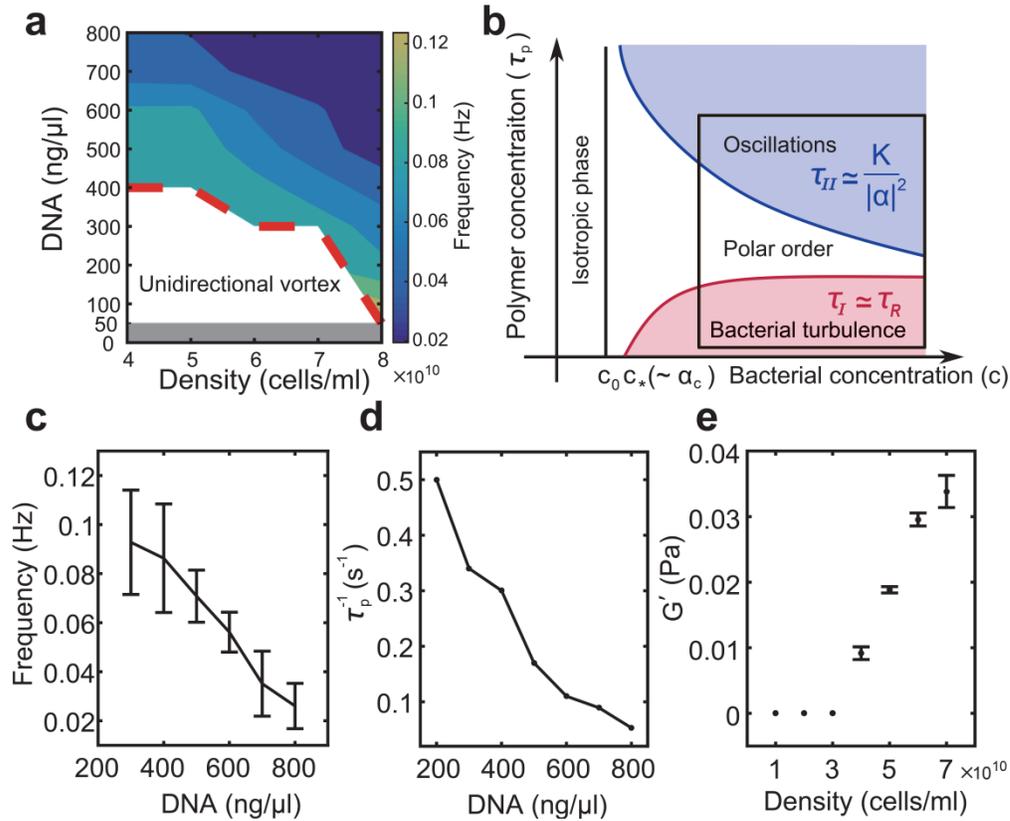

**Figure 4. Phase diagram of giant vortices and rheology measurement.** (**a**) Phase diagram from experiments. White area: unidirectional giant vortex; colored area, oscillatory giant vortex (colormap indicating chirality-switching frequency); grey area: bacterial turbulence. (**b**) Schematic phase diagram from linear instability analysis of the continuum model. We have taken the activity $|\alpha|$ and the Maxwell relaxation time $\tau_p$ as a proxy for bacterial and DNA concentrations, respectively. (**c**) Chirality-switching frequency in **a** versus DNA concentration at cell density $6 \times 10^{10}$ cells/mL. (**d**) Inverse of $\tau_p$ measured by microrheology as a function of DNA concentration (Methods). (**e**) Storage modulus of pure bacterial suspensions versus cell density (Methods; Extended Data Figure 9). Error bars in **c**,**e** indicate standard variation (N=5 and N=10, respectively).

**Extended Data Figures**

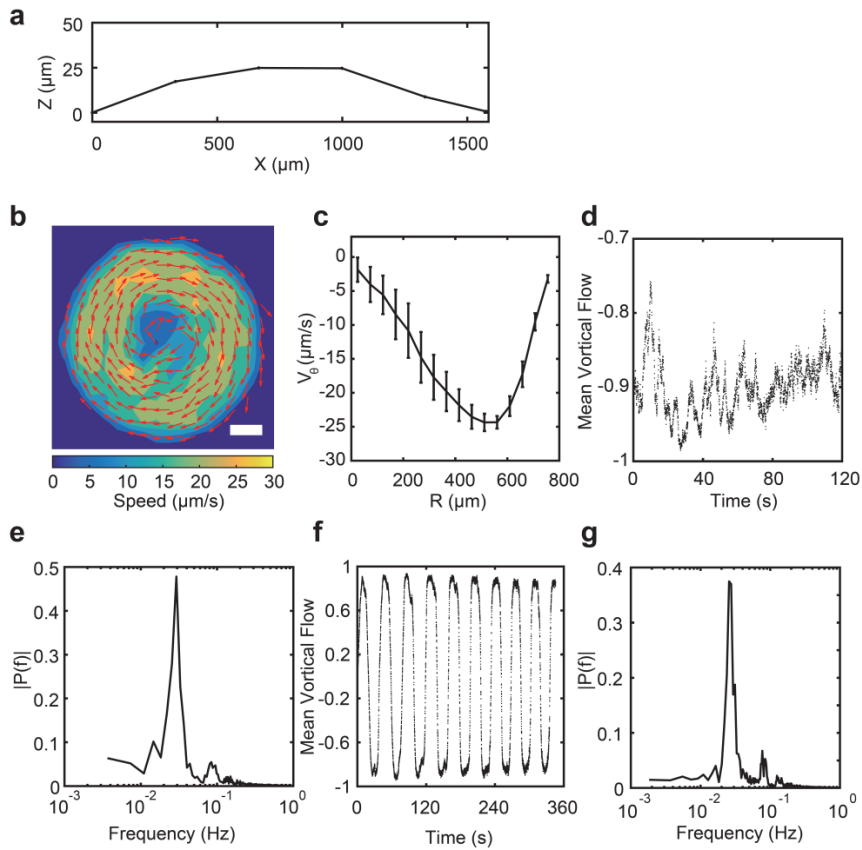

**Extended Data Figure 1. Height profile of bacterial suspension drop and further characterization of giant vortices.** (**a**) Height profile of a bacterial suspension drop measured by locating the uppermost and the lowermost focal planes where fluorescently-labelled cells could be found. (**b**) Instantaneous velocity field of a representative CW unidirectional giant vortex. Arrows and colormap represent collective velocity direction and magnitude, respectively (Methods). Cell density, $6 \times 10^{10}$ cells/mL; DNA concentration, 200 ng/µL. Scale bar, 250 µm. (**c**) Time- and azimuthally averaged tangential velocity $v_\theta(r)$ of the CW giant vortex in **b** plotted against radial position. Error bars represent standard deviation (N=1000 successive frames). (**d**) Normalized mean vortical flow of the CV giant vortex in **b** (Methods). (**e-g**) Fourier analysis of the normalized mean vortical flow ($P(t)$) in oscillatory giant vortices. (**e**) Fourier spectrum $|P(f)|$ for $P(t)$ in Figure 2b computed by Fast Fourier Transform, peaking at ~0.030 Hz and with a full width at half maximum (FWHM) of ~0.012 Hz. (**f**) $P(t)$ of an oscillatory vortex with 9 periods (cell density, ~6×10 cells/mL; DNA concentration, ~800 ng/µL). (**g**)

Fourier spectrum for $P(t)$ in panel **f**, peaking at ~0.026 Hz and with a FWHM of ~0.008 Hz.

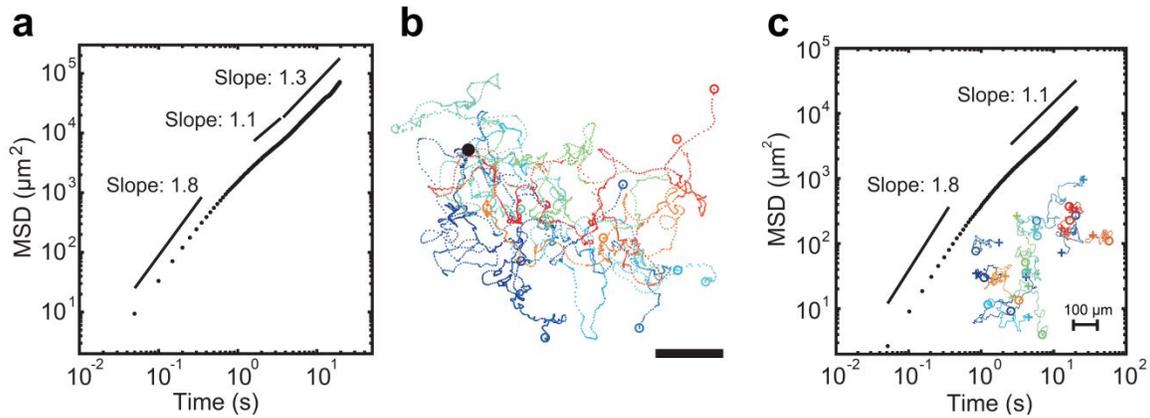

**Extended Data Figure 2. Diffusive behavior of single cells in giant vortices.** (**a**) The mean square displacement (MSD) of individual cells analyzed in main text Fig. 1e in the laboratory frame (Methods). (**b**) Bacterial trajectories in main text Fig. 1e replotted with the starting points shifted to the same position (black dot). Different color indicates different bacterium. Scale bar, 100 μm. (**c**) Local diffusive behavior of individual bacteria in an oscillatory giant vortex. MSD of cells was computed based on drift-subtracted single-cell trajectories. In a frame comoving with the giant vortex individual cells underwent ballistic motion at short time scale (~ 1 s) and diffusive motion over longer time scales. The diffusion constant $D$ was obtained by fitting the MSD at $t$ >2 s to $4Dt^{\alpha}$, yielding $D \approx 110 \ \mu\text{m}^2\text{s}^{-1}$ and $\alpha \approx 1.1$. In this oscillatory giant vortex, DNA concentration was ~500 ng/μL and cell density was ~6×10$^{10}$ cells/mL. Inset: Trajectories of 14 representative cells (+: starting point; ○: ending point). The time duration of each trajectory is ~28 s, about one period of the oscillatory giant vortex.

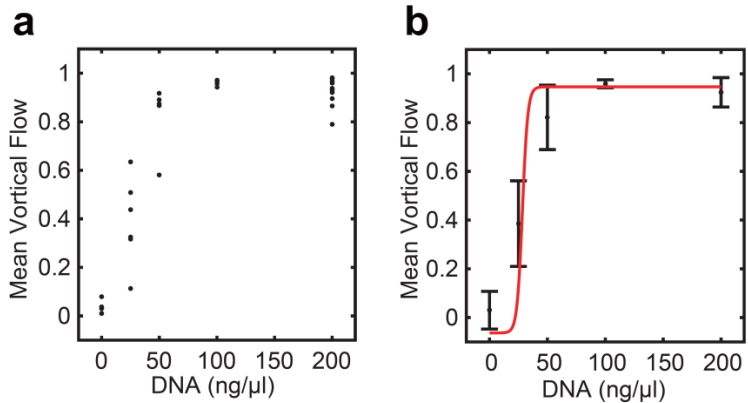

**Extended Data Figure 3.  Vortex order of bacterial suspension drop versus *E. coli* DNA concentration.**  The diameter of suspension drops were ~ 1.5 mm.  Cell density was fixed at 6×10 cells/mL.  (**a**) Scattered data points of vortex order (i.e. normalized mean vortical flow *P*) versus DNA concentration.  Each data point represents the normalized mean vortical flow averaged over a time window of ~20 s for one suspension drop with specific DNA concentration.  (**b**) Sigmoidal fit of *P* as a function of DNA concentration. The mean and standard deviation (error bars, N≥4) plotted in **b** were computed based on the scattered data points in **a**.  The data in **b** was fitted to a modified sigmoid function (Methods).

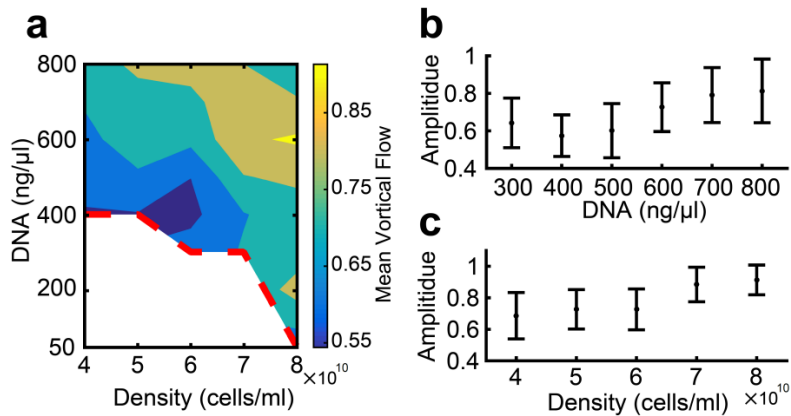

**Extended Data Figure 4.** Dependence of mean-vortical-flow amplitude of oscillatory giant vortices on cell density and DNA concentration. The mean-vortical-flow amplitude of a specific oscillatory giant vortex is taken as the averaged absolute value of extremums of the normalized mean vortical flow. (**a**) Contour plot of mean-vortical-flow amplitude (indicated by the colormap) in the plane of cell density and DNA concentration. Each data point in the contour plot is the average of mean-vortical-flow amplitude from at least 3 oscillatory giant vortices with the corresponding DNA concentration and cell density. (**b**) The mean-vortical-flow amplitude in panel **a** plotted against DNA concentration at fixed cell density ~6×10 cells/ml. (**c**) Mean-vortical-flow amplitude in panel **a** plotted against cell density at fixed DNA concentration ~600 ng/μL. Error bars in **b,c** indicate standard deviation (N≥3). Overall, the mean-vortical-flow amplitude of oscillatory giant vortices increases weakly with increasing DNA concentration and cell density.

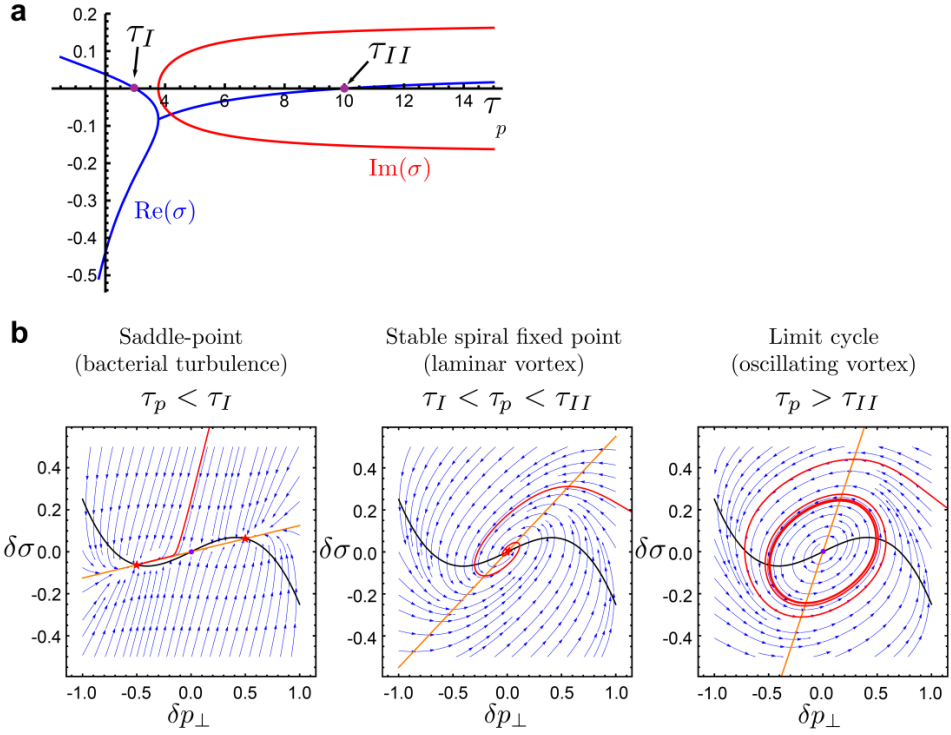

**Extended Data Figure 5.** (**a**) The mode structure as a function of $\tau_p$ for fixed $\alpha$, $\tau_R$ and $q \sim \sqrt{|\alpha|/K}$. For $\tau_p < \tau_I$, we have one purely real unstable mode ($\mathrm{Re}(\sigma) > 0$), while for $\tau_p > \tau_{II}$, the unstable modes have a finite frequency of oscillation. (**b**) The phase plane portraits in the $\{\delta p_\perp, \delta\sigma\}$ plane for the three different regimes- $\tau_p < \tau_I$, $\tau_I < \tau_p < \tau_{II}$, and $\tau_p > \tau_{II}$. We have included the leading gradient free nonlinear term $\delta p_\perp^3$ to saturate the polarization when unstable. This makes the system akin to the FitzHugh-Nagumo model for $\tau_p \gtrsim \tau_{II}$, leading to relaxation oscillations and excitability. The black and orange lines are the nullclines, and the red line is a representative trajectory that either converges to a fixed point or to a limit cycle. The red stars at the intersection of the nullclines are stable fixed points (or foci), while the blue dots are unstable fixed points (or foci). The labels to the three frames highlight the correspondence between the nature of the dynamical state obtained from the FitzHugh-Nagumo model and the states observed in experiments.

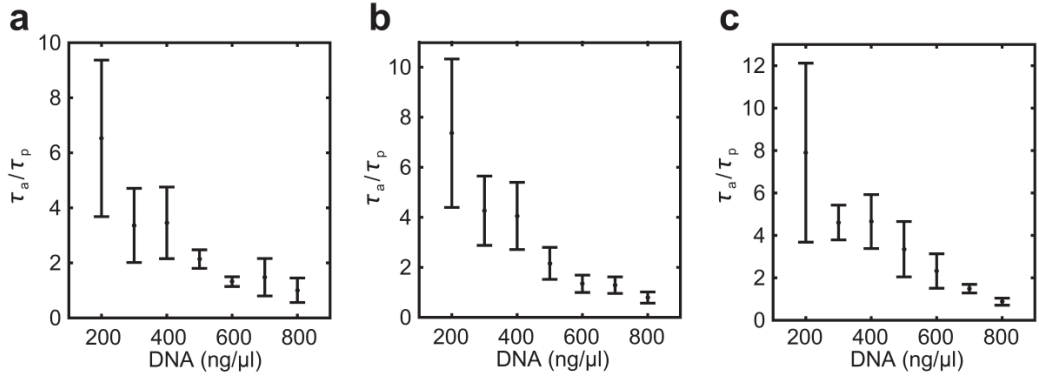

**Extended Data Figure 6. Ratio between active shearing time and DNA relaxation time in giant vortices plotted against DNA concentration.** The Maxwell relaxation time of DNA solutions $\tau_p$ was measured by microrheology (Methods). The active shearing time scale $\tau_a = \Gamma\,\ell_a^2/|\alpha| \sim K/\alpha^2$ in giant vortices cannot be computed precisely, since the relevant parameters are unknown. Instead, $\tau_a$ is estimated as the inverse of shear rate associated with bacterial collective motion, i.e. the correlation length of collective velocity field divided by mean collective speed. Cell density was fixed at 4×10, 6×10 and 8×10 cells/mL for panels **a**, **b** and **c**, respectively. The mean and uncertainty of each data point in the plots was computed based on the data of $\tau_p$ and $\tau_a$ measured from at least 3 giant vortices. Overall, $\tau_p$ approaches $\tau_a$ when unidirectional vortices transit to an oscillatory state, a result qualitatively consistent with our active matter model.

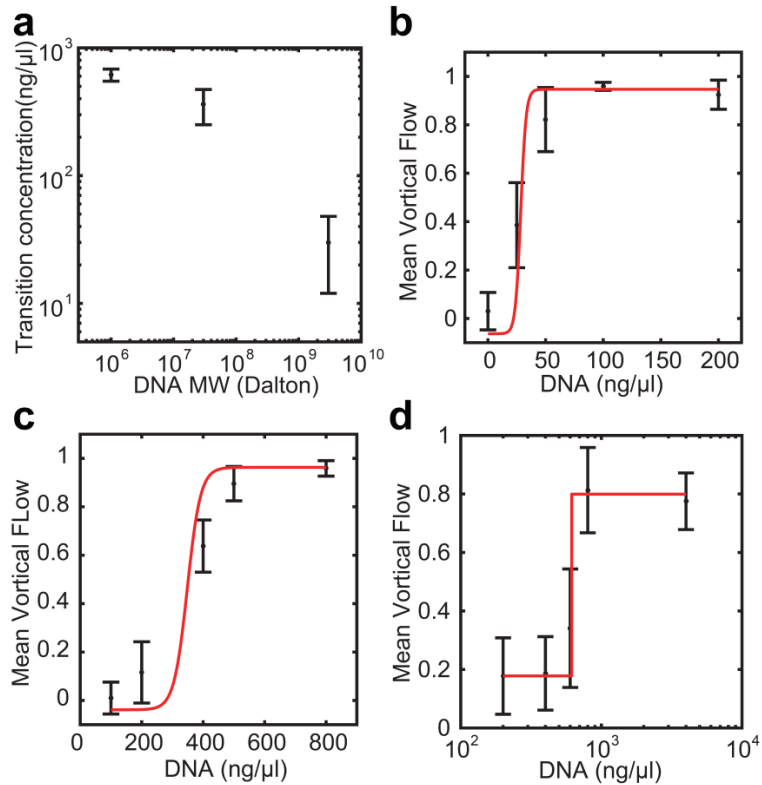

**Extended Data Figure 7. High molecular weight DNA can give rise to giant vortices**. (**a**) DNA concentration threshold ($d_1$) for the transition from bacterial turbulence to unidirectional giant vortex decreases with molecular weight (*N*) for the 3 types of DNA tested (*E. coli* genomic DNA, lambda phage DNA, and salmon testes DNA; Methods). The transition DNA concentration threshold and its uncertainty (indicated by error bars) for different types of DNA molecules was estimated based on sigmoidal fit of the normalized mean vortical flow as a function of DNA concentration; see Methods. The dependence of the threshold concentration $d_1$ appears to be consistent with the power-law scaling predicted for the effect of a dilute polymer solution on flow alignment and nematic viscosity [41]. (**b-d**) Normalized mean vortical flow of bacterial suspension drop versus *E. coli* DNA concentration obtained (**b**: *E. coli* genomic DNA, same as Extended Data Figure 3b and replotted here for comparison; **c**: lambda phage DNA; **d**: salmon testes DNA). Error bars indicate standard deviation, N≥4. The data in these plots were obtained in the same way as in Extended Data Figure 3b. The diameter of suspension drops were ~1.5 mm. Cell density was fixed at 6×10 cells/mL.

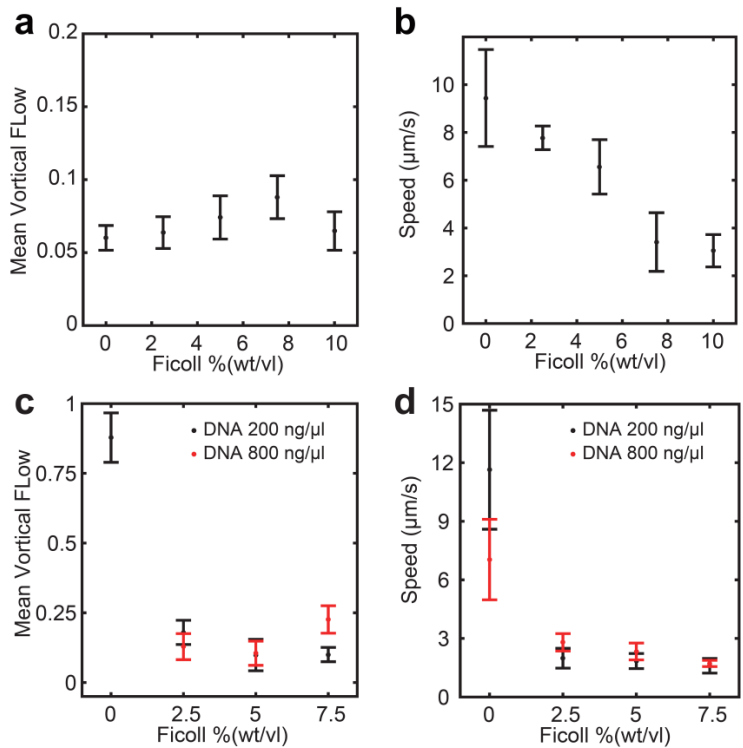

**Extended Data Figure 8. Effect of viscosity on bacterial collective motion in suspension drops.** The suspension drop diameter was ~ 1.5 mm. Cell density was fixed at 6×10 cells/mL. (**a,b**) Mean vortex order and average collective speed of bacterial suspension drops without additive DNA plotted against Ficoll (Ficoll 400, mol. wt. 400 kDa; Sigma, cat. No. F9378) concentration. The mean vortex order of a specific suspension drop was computed as the time average of absolute instantaneous vortex order (i.e. normalized mean vortical flow) over a time window of ~20 s. For a specific Ficoll concentration, the average collective speed of a suspension drop was computed as the time average of collective speed over a time window of ~20 s. (**c,d**) Mean vortex order and average collective speed of bacterial suspension drops with additive DNA plotted against Ficoll concentration. Black (or red) color indicates the experiments with *E. coli* genomic DNA concentration 200 (or 800) ng/μL, which normally supports the development of unidirectional or oscillatory giant vortices, respectively. Neither stable unidirectional giant vortex nor oscillatory giant vortex could be observed at all Ficoll concentrations (without additive DNA) or at Ficoll concentrations ≥2.5% (with DNA). Error bars in **a-d** indicate standard deviation (N≥5 suspension drops).

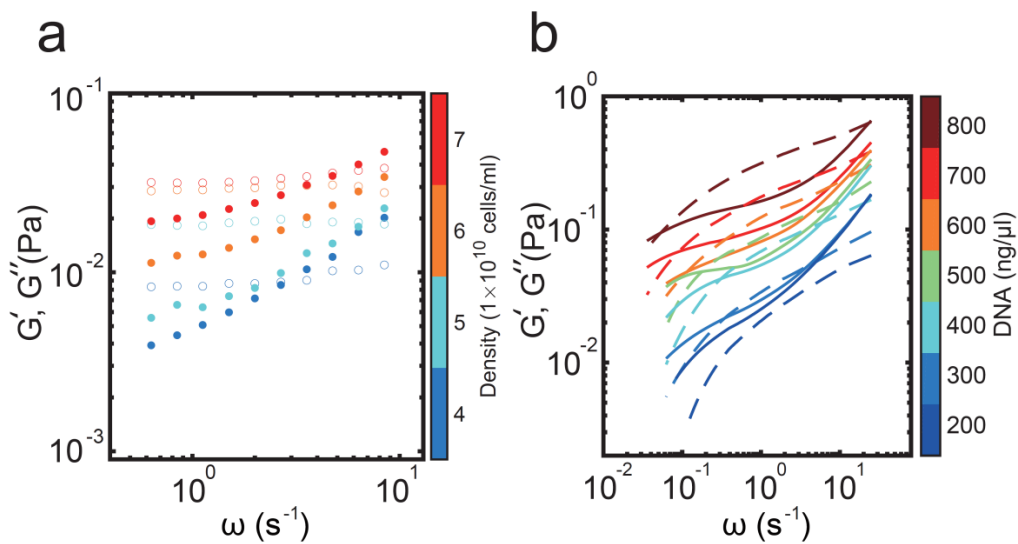

**Extended Data Figure 9. Dynamic modulus of pure bacterial suspension and DNA solution.** (**a**) Dynamic modulus of pure bacterial suspension measured by rheometer as a function of frequency (Methods), showing viscoelasticity consistent with Kelvin–Voigt model. The measurement was made on the scale of ~100 μm, comparable to the length scale of bacterial collective motion. Open circles represent storage modulus (*G′*); solid circles represent loss modulus (*G″*). Colormap indicates cell density. The elastic modulus of bacterial suspension measured in the range of ~ 0.1-1 Hz was used to compute data points in Fig. 4e in main text. (**b**) Dynamic modulus of DNA solution measured by microrheology (Methods). Dash line represents storage modulus (*G′*); solid line represents loss modulus (*G″*). Colormap indicates DNA concentration. The DNA solution behaves as Maxwell material. Note that the viscosity η of DNA solutions obtained from our microrheology measurement is much higher than that of water (e.g. η ~0.106 Pa·s at DNA concentration 200 ng/μL; Methods). The fact that cells were able to swim at a normal speed of ~20-30 μm/s at DNA concentrations we tested here suggest that swimming bacteria induce strong shear thinning effect in DNA solutions.

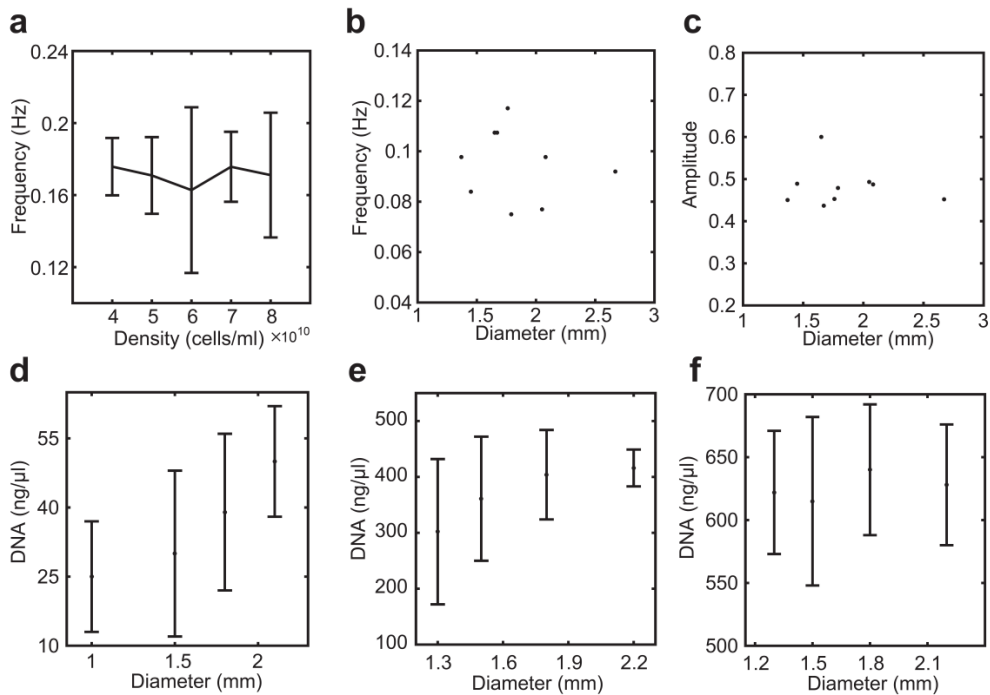

**Extended Data Figure 10. Confinement effect on the development of giant vortex state.** (**a**) Without spatial confinement (e.g. in centimeter-scale bacterial swarming colonies), dense bacterial active fluids can display collective oscillatory motion with the oscillation frequency independent of cell density as shown in the plot here (error bars indicate standard variation; N=5). (**b,c**) Oscillation frequency (panel **b**) and vortical flow amplitude (panel **c**) in oscillatory giant vortices plotted against confinement size (i.e. diameter of suspension drops). Each dot in **b,c** represents the data from one suspension drop with the specified size. Cell density was fixed at ~ 6×10 cells/mL and *E. coli* genomic DNA concentration was fixed at ~ 300 ng/µL. (**d-f**) DNA concentration threshold for the transition from bacterial turbulence to unidirectional giant vortex plotted against confinement size in the case of *E. coli* genomic DNA (panel **d**), lambda phage DNA (panel **e**) and salmon testes DNA (panel **f**). The DNA concentration threshold and its uncertainty (indicated by error bars) were estimated based on sigmoidal fit of normalized mean vortical flow as a function of DNA concentration (Methods). Cell density in **d-f** was fixed at ~ 6×10 cells/ml. Taken together, spatial confinement is necessary but not sufficient for giant vortex development.